\documentclass[12pt]{article}
\usepackage{graphicx,color}
\usepackage{amsmath}
\usepackage{XoohmE}
\usepackage{booktabs}

\def\D{{\rm D}}
\def\Dd{{\bar{\rm D}}}
\def\gd{{\dot{\gamma}}}

\def\E{{\mathcal{E}}}

\def\be{\begin{equation}}
\def\ee{\end{equation}}
\usepackage[retainorgcmds]{IEEEtrantools}

\def\bea{\begin{IEEEeqnarray*}}
\def\eea{\end{IEEEeqnarray*}}

\definecolor{Hey}{rgb}{.9,.05,.4}
\definecolor{orange}{rgb}{1,.5,0}
\definecolor{plum}{rgb}{.4,0,.6}
\definecolor{R}{rgb}{1,0,0}
\definecolor{G}{rgb}{0,1,0}
\definecolor{B}{rgb}{0,0,1}

\def\vCent#1{\vcenter{\hbox{\hss#1\hss}}}

\def\mathfrak{\bf}

\def\be{\begin{equation}}
\def\ee{\end{equation}}
\def\bea{\begin{eqnarray}}
\def\eea{\end{eqnarray}}

\def\dt#1{\on{\hbox{\bf .}}{#1}}                
\def\Dot#1{\dt{#1}}
\def\IR{\relax{\rm I\kern-.18em R}}
\def\binomial#1#2{\left(\,{\buildrel 
{\raise4pt\hbox{$\displaystyle{#1}$}}\over
{\raise-6pt\hbox{$\displaystyle{#2}$}}}\,\right)}

\def\[{\lfloor{\hskip 0.35pt}\!\!\!\lceil}
\def\]{\rfloor{\hskip 0.35pt}\!\!\!\rceil}

\catcode`@=11
\def\un#1{\relax\ifmmode\@@underline#1\else
        $\@@underline{\hbox{#1}}$\relax\fi}
\catcode`@=12

\def\ad{{\kern0.5pt
                   \alpha \kern-5.05pt
\raise5.8pt\hbox{$\textstyle.$}\kern
0.5pt}}

\def\Dot#1{{\kern0.5pt
     {#1} \kern-5.05pt \raise5.8pt\hbox{$\textstyle.$}\kern
0.5pt}}

\def\a{\alpha}
\def\b{\beta}
\def\c{\chi}
\def\d{\delta}

\def\g{\gamma}

\def\i{\iota}
\def\j{\psi}
\def\k{\kappa}
\def\l{\lambda}

\def\o{\omega}

\def\r{\rho}

\def\t{\tau}

\def\D{\Delta}

\def\L{\Lambda}
\def\O{\Omega}
\def\P{\Pi}

\def\S{\Sigma}

\def\ca{{\cal A}}

\def\cf{{\cal F}}

\def\bo{{\raise.15ex\hbox{\large$\Box$}}}               
\def\pa{\partial}                                       
\def\TH{{\raise.2ex\hbox{$\displaystyle \bigodot$}\mskip-4.7mu \llap H
\;}}
\def\face{{\raise.2ex\hbox{$\displaystyle \bigodot$}\mskip-2.2mu \llap
{$\ddot
        \smile$}}}                                      

\def\leftrightarrowfill{$\mathsurround=0pt \mathord\leftarrow \mkern-6mu
        \cleaders\hbox{$\mkern-2mu \mathord- \mkern-2mu$}\hfill
        \mkern-6mu \mathord\rightarrow$}
\def\dvec#1{\vbox{\ialign{##\crcr
        \leftrightarrowfill\crcr\noalign{\kern-1pt\nointerlineskip}
        $\hfil\displaystyle{#1}\hfil$\crcr}}}           
\def\dt#1{{\buildrel {\hbox{\LARGE .}} \over {#1}}}     

\def\frac#1#2{{\textstyle{#1\over\vphantom2\smash{\raise.20ex
        \hbox{$\scriptstyle{#2}$}}}}}                   
\def\sfrac#1#2{{\vphantom1\smash{\lower.5ex\hbox{\small$#1$}}\over
        \vphantom1\smash{\raise.4ex\hbox{\small$#2$}}}} 
\def\bfrac#1#2{{\vphantom1\smash{\lower.5ex\hbox{$#1$}}\over
        \vphantom1\smash{\raise.3ex\hbox{$#2$}}}}       
\def\afrac#1#2{{\vphantom1\smash{\lower.5ex\hbox{$#1$}}\over#2}}    

\newskip\humongous \humongous=0pt plus 1000pt minus 1000pt
\def\caja{\mathsurround=0pt}
\def\eqalign#1{\,\vcenter{\openup2\jot \caja
        \ialign{\strut \hfil$\displaystyle{##}$&$
        \displaystyle{{}##}$\hfil\crcr#1\crcr}}\,}
\newif\ifdtup
\def\panorama{\global\dtuptrue \openup2\jot \caja
        \everycr{\noalign{\ifdtup \global\dtupfalse
        \vskip-\lineskiplimit \vskip\normallineskiplimit
        \else \penalty\interdisplaylinepenalty \fi}}}

  \def\pp{{\mathchoice
          {
              \kern 1pt%
              \raise 1pt
              \vbox{\hrule width5pt height0.4pt depth0pt
                    \kern -2pt
                    \hbox{\kern 2.3pt
                          \vrule width0.4pt height6pt depth0pt
                          }
                    \kern -2pt
                    \hrule width5pt height0.4pt depth0pt}%
                    \kern 1pt
           }
            {
              \kern 1pt%
              \raise 1pt
              \vbox{\hrule width4.3pt height0.4pt depth0pt
                    \kern -1.8pt
                    \hbox{\kern 1.95pt
                          \vrule width0.4pt height5.4pt depth0pt
                          }
                    \kern -1.8pt
                    \hrule width4.3pt height0.4pt depth0pt}%
                    \kern 1pt
            }
            {
              \kern 0.5pt%
              \raise 1pt
              \vbox{\hrule width4.0pt height0.3pt depth0pt
                    \kern -1.9pt  
                    \hbox{\kern 1.85pt
                          \vrule width0.3pt height5.7pt depth0pt
                          }
                    \kern -1.9pt
                    \hrule width4.0pt height0.3pt depth0pt}%
                    \kern 0.5pt
            }
            {
              \kern 0.5pt%
              \raise 1pt
              \vbox{\hrule width3.6pt height0.3pt depth0pt
                    \kern -1.5pt
                    \hbox{\kern 1.65pt
                          \vrule width0.3pt height4.5pt depth0pt
                          }
                    \kern -1.5pt
                    \hrule width3.6pt height0.3pt depth0pt}%
                    \kern 0.5pt
            }
        }}

  \def\mm{{\mathchoice
                       {
                             \kern 1pt
               \raise 1pt    \vbox{\hrule width5pt height0.4pt depth0pt
                                  \kern 2pt
                                  \hrule width5pt height0.4pt depth0pt}
                             \kern 1pt}
                       {
                            \kern 1pt
               \raise 1pt \vbox{\hrule width4.3pt height0.4pt depth0pt
                                  \kern 1.8pt
                                  \hrule width4.3pt height0.4pt depth0pt}
                             \kern 1pt}
                       {
                            \kern 0.5pt
               \raise 1pt
                            \vbox{\hrule width4.0pt height0.3pt depth0pt
                                  \kern 1.9pt
                                  \hrule width4.0pt height0.3pt depth0pt}
                            \kern 1pt}
                       {
                           \kern 0.5pt
             \raise 1pt  \vbox{\hrule width3.6pt height0.3pt depth0pt
                                  \kern 1.5pt
                                  \hrule width3.6pt height0.3pt depth0pt}
                           \kern 0.5pt}
                       }}

\def\pd{{\kern0.5pt
                   + \kern-5.05pt \raise5.8pt\hbox{$\textstyle.$}\kern
0.5pt}}

\def\pmd{{\kern0.5pt
                  \pm \kern-5.05pt \raise6.3pt\hbox{$\textstyle.$}\kern1.5pt}}

\def\md{{\mathchoice
   {
      {{\kern 1pt - \kern-6.2pt \raise5pt\hbox{$\textstyle.$}\kern 1pt}}}
    {
      {{\kern 1pt - \kern-6.2pt \raise5pt\hbox{$\textstyle.$}\kern 1pt}}}
    {
      {\kern0.5pt - \kern-5.05pt \raise3.4pt\hbox{$\textstyle.$}\kern0.5pt}}
    {
      {\kern0.5pt - \kern-5.05pt \raise3.4pt\hbox{$\textstyle.$}\kern0.5pt}}}}

\def\ad{{\dot{\alpha}}}
\def\bd{{\dot{\beta}}}

\def\pp{{\mathchoice
          {
              \kern 1pt%
              \raise 1pt
              \vbox{\hrule width5pt height0.4pt depth0pt
                    \kern -2pt
                    \hbox{\kern 2.3pt
                          \vrule width0.4pt height6pt depth0pt
                          }
                    \kern -2pt
                    \hrule width5pt height0.4pt depth0pt}%
                    \kern 1pt
           }
            {
              \kern 1pt%
              \raise 1pt
              \vbox{\hrule width4.3pt height0.4pt depth0pt
                    \kern -1.8pt
                    \hbox{\kern 1.95pt
                          \vrule width0.4pt height5.4pt depth0pt
                          }
                    \kern -1.8pt
                    \hrule width4.3pt height0.4pt depth0pt}%
                    \kern 1pt
            }
            {
              \kern 0.5pt%
              \raise 1pt
              \vbox{\hrule width4.0pt height0.3pt depth0pt
                    \kern -1.9pt  
                    \hbox{\kern 1.85pt
                          \vrule width0.3pt height5.7pt depth0pt
                          }
                    \kern -1.9pt
                    \hrule width4.0pt height0.3pt depth0pt}%
                    \kern 0.5pt
            }
            {
              \kern 0.5pt%
              \raise 1pt
              \vbox{\hrule width3.6pt height0.3pt depth0pt
                    \kern -1.5pt
                    \hbox{\kern 1.65pt
                          \vrule width0.3pt height4.5pt depth0pt
                          }
                    \kern -1.5pt
                    \hrule width3.6pt height0.3pt depth0pt}%
                    \kern 0.5pt
            }
        }}

  \def\mm{{\mathchoice
                       {
                             \kern 1pt
               \raise 1pt    \vbox{\hrule width5pt height0.4pt depth0pt
                                  \kern 2pt
                                  \hrule width5pt height0.4pt depth0pt}
                             \kern 1pt}
                       {
                            \kern 1pt
               \raise 1pt \vbox{\hrule width4.3pt height0.4pt depth0pt
                                  \kern 1.8pt
                                  \hrule width4.3pt height0.4pt depth0pt}
                             \kern 1pt}
                       {
                            \kern 0.5pt
               \raise 1pt
                            \vbox{\hrule width4.0pt height0.3pt depth0pt
                                  \kern 1.9pt
                                  \hrule width4.0pt height0.3pt depth0pt}
                            \kern 1pt}
                       {
                           \kern 0.5pt
             \raise 1pt  \vbox{\hrule width3.6pt height0.3pt depth0pt
                                  \kern 1.5pt
                                  \hrule width3.6pt height0.3pt depth0pt}
                           \kern 0.5pt}
                       }}

\def\pd{{\kern0.5pt
                   + \kern-5.05pt \raise5.8pt\hbox{$\textstyle.$}\kern
0.5pt}}

\def\pmd{{\kern0.5pt
                  \pm \kern-5.05pt \raise6.3pt\hbox{$\textstyle.$}\kern1.5pt}}

\def\md{{\mathchoice
   {
      {{\kern 1pt - \kern-6.2pt \raise5pt\hbox{$\textstyle.$}\kern 1pt}}}
    {
      {{\kern 1pt - \kern-6.2pt \raise5pt\hbox{$\textstyle.$}\kern 1pt}}}
    {
      {\kern0.5pt - \kern-5.05pt \raise3.4pt\hbox{$\textstyle.$}\kern0.5pt}}
    {
      {\kern0.5pt - \kern-5.05pt \raise3.4pt\hbox{$\textstyle.$}\kern0.5pt}}}}

\def\dslash{\not{\hbox{\kern-2pt $\partial$}}}
\def\Dslash{\not{\hbox{\kern-4pt $D$}}}
\def\pslash{\not{\hbox{\kern-2.3pt $p$}}}
 \newtoks\slashfraction
 \slashfraction={.13}
 \def\slash#1{\setbox0\hbox{$ #1 $}
 \setbox0\hbox to \the\slashfraction\wd0{\hss \box0}/\box0 }
 
\font\ro=cmsy10                          
\def\kcr{{\hbox{\ro \char'170}}}                
\def\ktl{{\hbox{\ro \char'170}}}        
\def\ktr{{\hbox{\ro \char'170}}}        
\def\kbl{{\hbox{\ro \char'170}}}        
\def\kbr{{\hbox{\ro \char'170}}}        

\def\plpl{\raise-2pt\hbox{$\raise3pt\hbox{$_+$}\hskip-6.67pt\raise0.0pt
\hbox{$^+$}\hskip 0.01pt$}}
\def\mimi{\raise-2pt\hbox{$\raise3pt\hbox{$_-$}\hskip-6.67pt\raise0.0pt
\hbox{$^-$}\hskip 0.01pt$}} 

\def\bo{{\raise.15ex\hbox{\large$\Box$}}}               
\def\pa{\partial}                                       
\def\TH{{\raise.2ex\hbox{$\displaystyle \bigodot$}\mskip-4.7mu \llap H \;}}
\def\face{{\raise.2ex\hbox{$\displaystyle \bigodot$}\mskip-2.2mu \llap {$\ddot
        \smile$}}}                                      

\def\leftrightarrowfill{$\mathsurround=0pt \mathord\leftarrow \mkern-6mu
        \cleaders\hbox{$\mkern-2mu \mathord- \mkern-2mu$}\hfill
        \mkern-6mu \mathord\rightarrow$}
\def\dvec#1{\vbox{\ialign{##\crcr
        \leftrightarrowfill\crcr\noalign{\kern-1pt\nointerlineskip}
        $\hfil\displaystyle{#1}\hfil$\crcr}}}           
\def\dt#1{{\buildrel {\hbox{\LARGE .}} \over {#1}}}     

\def\frac#1#2{{\textstyle{#1\over\vphantom2\smash{\raise.20ex
        \hbox{$\scriptstyle{#2}$}}}}}                   
\def\sfrac#1#2{{\vphantom1\smash{\lower.5ex\hbox{\small$#1$}}\over
        \vphantom1\smash{\raise.4ex\hbox{\small$#2$}}}} 
\def\bfrac#1#2{{\vphantom1\smash{\lower.5ex\hbox{$#1$}}\over
        \vphantom1\smash{\raise.3ex\hbox{$#2$}}}}       
\def\afrac#1#2{{\vphantom1\smash{\lower.5ex\hbox{$#1$}}\over#2}}    

\parskip=\medskipamount                 
\thicklines                         

\def\oldheadpic{                                
        \setlength{\unitlength}{.4mm}
        \thinlines
        \par
        \begin{picture}(349,16)
        \put(325,16){\line(1,0){4}}
        \put(330,16){\line(1,0){4}}
        \put(340,16){\line(1,0){4}}
        \put(335,0){\line(1,0){4}}
        \put(340,0){\line(1,0){4}}
        \put(345,0){\line(1,0){4}}
        \put(329,0){\line(0,1){16}}
        \put(330,0){\line(0,1){16}}
        \put(339,0){\line(0,1){16}}
        \put(340,0){\line(0,1){16}}
        \put(344,0){\line(0,1){16}}
        \put(345,0){\line(0,1){16}}
        \put(329,16){\oval(8,32)[bl]}
        \put(330,16){\oval(8,32)[br]}
        \put(339,0){\oval(8,32)[tl]}
        \put(345,0){\oval(8,32)[tr]}
        \end{picture}
        \par
        \thicklines
        \vskip.2in}
\def\oldtitle#1#2#3#4{\oldheadpic\begin{center}\vglue.5in{\large\bf #1}\\[.6in]
        {#2}\\[.1in] {\it Department of Physics and Astronomy}\\
        {\it University of Maryland, College Park, MD 20742}\\[.6in]
        Physics Publication \#{#3}\\ {#4}\\[1.5in] {\bf ABSTRACT}\\[.1in]
        \end{center} \begin{quotation}}                 
\def\oldTitle#1#2#3#4#5#6#7{\oldheadpic\begin{center} \vglue .4in
        {\large\bf #1}\\[.4in]
        {#2}\\[.1in] {\it Department of Physics and Astronomy}\\
        {\it University of Maryland, College Park, MD 20742}\\[.1in]
        {#3}\\[.1in] {\it {#4}}\\ {\it {#5}}\\[.4in]
        Physics Publication \#{#6}\\ {#7}\\[.5in] {\bf ABSTRACT}\\[.1in]
        \end{center} \begin{quotation}}                 
\def\border{                                            
        \setlength{\unitlength}{1mm}
        \newcount\xco
        \newcount\yco
        \xco=-21
        \yco=12
        \begin{picture}(140,0)
        \put(\xco,\yco){$\ktl$}
        \advance\yco by-1
        {\loop
        \put(\xco,\yco){$\kcr$}
        \advance\yco by-2
        \ifnum\yco>-240
        \repeat
        \put(\xco,\yco){$\kbl$}}
        \xco=158
        \yco=12
        \put(\xco,\yco){$\ktr$}
        \advance\yco by-1
        {\loop
        \put(\xco,\yco){$\kcr$}
        \advance\yco by-2
        \ifnum\yco>-240
        \repeat
        \put(\xco,\yco){$\kbr$}}
        \put(-20,13){\tiny **University of Maryland * Center for String and 
         Particle  Theory* Physics Department***University of Maryland *Center  
        for String and Particle  Theory** }
        \put(-20,-241.5){\tiny **University of Maryland * Center for String and 
         Particle  Theory* Physics Department***University of Maryland *Center  
        for String and Particle  Theory** }
        \end{picture}
        \par\vskip-8mm}
\def\bordero{                                           
        \setlength{\unitlength}{1mm}
        \newcount\xco
        \newcount\yco
        \xco=-31
        \yco=12
        \begin{picture}(140,0)
        \put(\xco,\yco){$\ktl$}
        \advance\yco by-1
        {\loop
        \put(\xco,\yco){$\kclr}
        \advance\yco by-2
        \ifnum\yco>-240
        \repeat
        \put(\xco,\yco){$\kbl$}}
        \xco=151
        \yco=12
        \put(\xco,\yco){$\ktr$}
        \advance\yco by-1
        {\loop
        \put(\xco,\yco){$\kcr$}
        \advance\yco by-2
        \ifnum\yco>-240
        \repeat
        \put(\xco,\yco){$\kbr$}}
        \put(-20,12){\ooo bacdefghidfghghdhededbihdgdfdfhhdheidhdhebaaahjhhdahba

hgdedge
   hgfdiehhgdigicba}
        \put(-20,-241.5){\ooo ababaighefdbfghgeahgdfgafagihdidihiidhiagfedhadbfd

ecdcdfa
   gdcbhaddhbgfchbgfdacfediacbabab}
        \end{picture}
        \par\vskip-8mm}
\def\headpic{                                           
        \indent
        \setlength{\unitlength}{.4mm}
        \thinlines
        \par
        \begin{picture}(29,16)
        \put(165,16){\line(1,0){4}}
        \put(170,16){\line(1,0){4}}
        \put(180,16){\line(1,0){4}}
        \put(175,0){\line(1,0){4}}
        \put(180,0){\line(1,0){4}}
        \put(185,0){\line(1,0){4}}
        \put(169,0){\line(0,1){16}}
        \put(170,0){\line(0,1){16}}
        \put(179,0){\line(0,1){16}}
        \put(180,0){\line(0,1){16}}
        \put(184,0){\line(0,1){16}}
        \put(185,0){\line(0,1){16}}
        \put(169,16){\oval(8,32)[bl]}
        \put(170,16){\oval(8,32)[br]}
        \put(179,0){\oval(8,32)[tl]}
        \put(185,0){\oval(8,32)[tr]}
        \end{picture}
        \par\vskip-6.5mm
        \thicklines}
\def\title#1#2#3#4{\border\headpic {\hbox to\hsize{#4 \hfill UMDEPP #3}}\par
        \begin{center} \vglue .5in {\large\bf #1}\\[.6in]
        {#2}\\[.1in] {\it Department of Physics and Astronomy}\\
        {\it University of Maryland, College Park, MD 20742}\\[1.5in]
        {\bf ABSTRACT}\\[.1in] \end{center} \begin{quotation}}  
\def\Title#1#2#3#4#5#6#7{\border\headpic
        {\hbox to\hsize{#7 \hfill UMDEPP #6}}\par
        \begin{center} \vglue .4in {\large\bf #1}\\[.4in]
        {#2}\\[.1in] {\it Department of Physics and Astronomy}\\
        {\it University of Maryland, College Park, MD 20742}\\[.1in]
        {#3}\\[.1in] {\it {#4}}\\ {\it {#5}}\\[.5in] {\bf ABSTRACT}\\[.1in]
        \end{center} \begin{quotation}}                 
\def\endtitle{\end{quotation}\newpage}                  

\def\qd{{\kern0.5pt
                   q \kern-5.05pt \raise5.8pt\hbox{$\textstyle.$}\kern
0.5pt}}

\begin{document}

\border\headpic {\hbox to\hsize{October 2013 \hfill
{UMDEPP-013-017}}}
\par
{$~$ \hfill
}
\par

\setlength{\oddsidemargin}{0.3in}
\setlength{\evensidemargin}{-0.3in}
\begin{center}
\vglue .10in
{\large\bf A Dynamical Theory for Massive Supergravity\footnote
{Supported in part  by National Science Foundation Grant
PHY-09-68854.}\  }
\\[.5in]

S.\, James Gates, Jr.\footnote{gatess@wam.umd.edu}
and
Konstantinos Koutrolikos\footnote{koutrol@umd.edu}
\\[0.2in]

{\it Center for String and Particle Theory\\
Department of Physics, University of Maryland\\
College Park, MD 20742-4111 USA}\\[3.6in]

{\bf ABSTRACT}\\[.01in]
\end{center}
\begin{quotation}
{
We present a new massive theory of superspin $Y=3/2$ which 
has non-minimal supergravity as it's massless limit. The new 
result will illuminate the underlying structure of auxiliary fields 
required for the description of arbitrary massive half-integer 
superspin systems.
}

\endtitle

\section{Introduction}
~~ After four decades of exploring the topic of supersymmetry
(SUSY), the problem of writing a manifestly susy-invariant action 
that describes a free, off-shell massive arbitrary superspin irreducible 
representation of the Super-Poincar\'{e} group still possesses 
puzzles. Although the non-supersymmetric case of massive higher 
spin theory has been developed~\cite{Singh:1974qz},\cite{Singh:1974rc} 
and is well understood, the off-shell supersymmetric case has yet 
to be understood with a comparable level of clarity.  There has 
been progress for on-shell supersymmetry ~\cite{Zinoviev:2007js}, 
but these results don't capture the rich off-shell structure of supersymmetric 
theories. There is a need for a manifestly susy invariant theory 
of massive integer and half-integer superspins which includes 
all the auxiliary superfields a theory of this nature is expected to possess.

Progress in this direction was made with the works 
presented in \cite{Buchbinder:2002gh},\cite{Buchbinder:2002tt},
\cite{Gates:2006cq}. These results provided a proof of concept that 
constructions like these are possible, but in these cases, the results 
results don't shed light to the heart of the problem 
which is to determine the set of auxiliary superfields required 
to describe an arbitrary superspin system with a proper 
massless limit.  Specifically in \cite{Buchbinder:2002gh} 
the focus was on massive extension of theories such as 
old-minimal supergravity, new-minimal supergravity, whose 
massless limits don't generalize to the arbitrary spin case. 
Therefore they do not provide clues about the underlying 
structure of auxiliary superfields for the arbitrary superspin
case.
 
This is not the case with the work presented in \cite{Gates:2006cq} 
where a massive extension of non-minimal supergravity is 
derived. The massless limit of that theory is non-minimal 
supergravity which is a member of an arbitrary super-helicity 
tower and that makes it a good starting point.  However the
derivation used a lagrange multiplier technique in order to 
impose constraints that were not be derived in a dynamical way.

We will show in the following that there is an alternative formulation 
of the theory where all the superfields are dynamical and the desired 
constraints follow from the equations of motion of these superfields. 
For this to work we require the presence of two fermionic auxiliary 
superfields. In the massless limit one of these decouples and the 
other one will play the role of the compensator in non-minimal 
supergravity. 

Our presentation is organized as follows: In section \ref{MassiveIrreps}, 
we quickly review the representation theory of the Super-Poincar\'{e} 
group for a massive half-integer superspin system. In section 
\ref{MasslessLimit}, we present the constraints imposed in the 
theory in order to have a proper massless limit. In the following 
section \ref{LowSpinExamples}, we start with a warm up exercise 
by quickly reproducing the massive theory for superspin $Y=1/2$. 
In the last section \ref{3/2} we present the new massive theory 
for $Y=3/2$.

\section{Half-Integer Superspin Representation Theory}
\label{MassiveIrreps}
~~ The irreducible representations of the Super-Poincar\'{e} group 
are labeled by it's two Casimir operators. The first one is the mass 
and the other one is a supersymmetric extension of the Poincar\'{e} 
Spin operator. For the massive case the Super-spin casimir operator 
takes the form
\be
C_2=\frac{W^2}{m^2}+\left(\frac{3}{4}+\lambda\right)P_{(o)}
~~~,
\ee
where $W^2$ is the ordinary spin operator, $P_{(o)}$ is a projection 
operator and the parameter $\lambda$ satisfies the equation
\be
\lambda^2+\lambda=\frac{W^2}{m^2} ~~~.
\ee

In order to diagonalize $C_2$ we want to diagonalize both $W^2,\ 
P_{(o)}$. The superfield $\Phi_{\a(n)\ad(m)}$ that does this and 
describes the highest possible representation (highest superspin)
\be
C_2\Phi_{\a(n)\ad(m)}=Y(Y+1)\Phi_{\a(n)\ad(m)},~Y=\frac{n+m+1}{2}
~~~,
\ee
has to satisfy the following:
\be \eqalign{
\D^2\Phi_{\a(n)\ad(m)}\,&=~ 0 ~~,  \cr 
\Dd^2\Phi_{\a(n)\ad(m)} \,&=~ 0 ~~,  \cr
\D^{\g}\Phi_{\g\a(n-1)\ad(m)} \,&=~ 0 ~~~, \cr
\pa^{\g\gd}\Phi_{\g\a(n-1)\gd\ad(m-1)} \,&=~ 0 ~~~, \cr
\Box\Phi_{\a(n)\ad(m)} \,&=~ m^2\Phi_{\a(n)\ad(m)} ~~~,
} \ee
where all dotted and undotted indices are fully symmetrized
and the spin content of this supermultiplet is $j=Y+1/2,\ Y,\ Y,\ Y-1/2$.

A superfield that describes a superspin $Y$ system has index structure 
such that $n+m=2Y-1$ where $n,m$ are integers. This Diophantine 
equation has a finite number of different solutions for $(n,m)$ pairs but 
the corresponding superfields are all equivalent because we can use 
the $\pa_{\b\bd}$ operator to convert one kind of index to another. So 
we can pick one of them to represent the entire class.

One last comment has to be made about the reality of the representation. 
The reality condition imposed on the superfield differs with the character 
of the superfield. For bosonic ones (even total number of indices), we can 
pick them to have $n=m$ and the reality condition is $\Phi_{\a(n)\ad(n)}=
\bar{\Phi}_{\a(n)\ad(n)}$. For fermionic superfields (odd total number of 
indices) we can pick $n=m+1$ and the reality condition is the Dirac 
equation $i\pa_{\a_n}{}^{\ad_n}\bar{\Phi}_{\a(n-1)\ad(n)}+m\Phi_{\a(n
)\ad(n-1)}$.

\section{The Massless Limit}
\label{MasslessLimit}

~~ Representation theory tells us the type of superfield and constraints 
we need in order to describe a specific irreducible representation. We 
would like to have a dynamical way to derive these constraints, through 
a lagrangian. That means we need a set of auxiliary superfields to help us 
generate these constraints. This is the core of the problem, to find the 
set of auxiliary fields and their interactions that accomplish these goals. 
That sounds like an intuitive trial-and-error process, but there is a hidden 
clue and...the massless limit of the theory.

As was illustrated in \cite{Kuzenko:1993jq}, \cite{Kuzenko:1993jp}, 
\cite{H-IntSpinI}, \cite{IntSpin}, and \cite{H-IntSpinII} there is one infinite 
tower for theories of integer superhelicity and two different infinite towers 
for theories of half integer super-helicities.

$$
\vCent
{\setlength{\unitlength}{1mm}
\begin{picture}(-20,0)
\put(-80,-82){\includegraphics[width=6.4in]{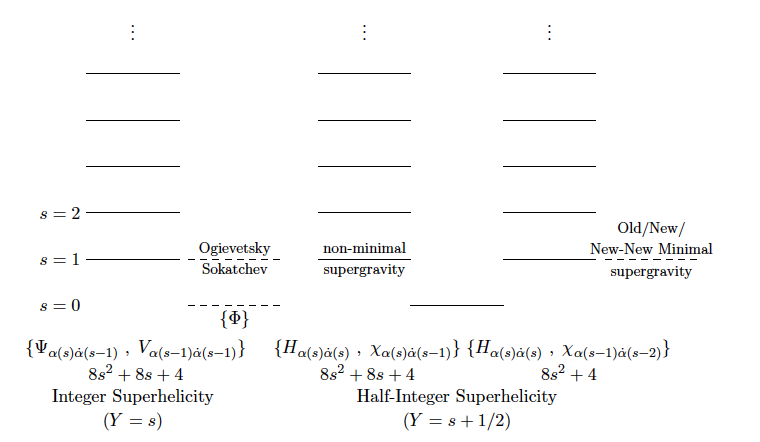}}
\put(-60,-90){Figure One: Towers of Massive Higher Spin Supermultiplets}
\end{picture}}
\nonumber
$$
\vskip3.4in
These theories were constructed under the requirement that the 
massless limit of a massive superspin $Y$ theory will give the 
massless theory of superhelicity $Y$ plus things that decouple. 
Therefore now that we want to build the massive theory we know 
what its massless limit must be. The conclusion is that the 
construction of the massive theories must start with the massless 
action and the addition of (self)interaction terms proportional to 
$m$ and $m^2$, so the massless theory decouples in the 
massless limit. Hence immediately and for free we obtain the 
first auxiliary field required. The massless theories are formulated 
in terms of a main superfield and a compensator. For the massive 
extension the compensator will become the first auxiliary superfield 
needed.

\section{Warming up with $Y=1/2$}
\label{LowSpinExamples}
~~ So if we want to construct the theory of superspin $1/2$ we start 
with the theory of superhelicity $1/2$, add terms proportional to 
$m$ and $m^2$ and check if we can generate the desired 
constraints. If not then we add extra auxiliary fields until we do. 
The starting action is:
\be  \eqalign{  {~~~~~~}
S=\int d^8z {\Big \{} ~&a_1 \, H{\rm D}^{\g}\Dd^2{\rm D}_{\g}H 
~+~ a_2m~H\left({\rm D}^2H+\Dd^2H\right)  
~+~ a_3m^2~H^2  {\Big \}} ~~~.
} \ee
To describe $Y=\frac{1}{2}$, $H$ must satisfy ${\rm D}^2H=0$ and 
$\Box H=m^2H$.  The equation of motion is
\be
\E^{(H)}=\frac{\d S}{\d H}
=2a_1{\rm D}^{\g}\Dd^2{\rm D}_{\g}H+2a_2m\left({\rm D}^2H+\Dd^2
H\right)+2a_3m^2H ~~~,
\ee
which gives
\be
{\rm D}^2\E^{(H)}=2a_2m {\rm D}^2\Dd^2H+2a_3m^2{\rm D}^2H ~~~,
\ee
so by choosing $a_2=0,a_3\neq 0$ we find ${\rm D}^2H=0$ $\leadsto 
\Dd^2H=0 ~ \text{(reality)}$ and if this is 
substituted back into $\E^{(H)}$ we get
$\Box H=\frac{a_3}{a_1}m^2H$ which fixes $a_3=a_1$ for compatibility 
with the Klein-Gordon equation.

There is also another way to obtain these results and that is \`{a} la St\"{u}ckelberg. 
The observation is that at least on-shell the massive superspin $\frac{1}{2}$ 
can be seen as the result of the combination of the massless superhelicity 
$\frac{1}{2}$ plus the massless superhelicity $0$. So we start with the actions 
for superhelicity $\frac{1}{2}$ and $0$ and we introduce (self)interaction terms 
proportional to $m$ and $m^2$
\be \eqalign{
S=\int d^8z & {\Big \{} a_1~H{\Dd}^{\gd} {\rm D}^2\Dd_{\gd}H   
+a_2m \, H\left({\rm D}^2H+\Dd^2H\right) +a_3m^2~H^2   \cr
&~~+\g m~H\left(\Phi+\bar{\Phi}\right)
~+\ b_1~\Phi\bar{\Phi} {\Big \} } ~+~
\int d^6z~ b_2m~\Phi\Phi  ~~~.
} \ee
The equations of motion are:
\be
\E^{(H)} = 2a_1{\rm D}^{\g}\Dd^2{\rm D}_{\g}H+2a_2m\left({\rm D}^2H+
\Dd^2H\right)+\g m\left(\Phi+\bar{\Phi}\right)+2a_3m^2H  ~~~, 
\ee
\be
\E^{(\Phi)}=-b_1\Dd^2\bar{\Phi}-\g m\Dd^2H+2b_2m\Phi  ~~~.
\ee
If we manage to show that on-shell $\Phi=0$ then $\E^{(\Phi)}=0\leadsto{
\rm D}^2H=0\leadsto\Box H=m^2H~(a_3=a_1)$.  With that goal in mind 
we attempt to eliminate $H$ from the equation of $\Phi$ and choose 
coefficients in such a way to find $\Phi=0$.  We begin by defining $I=
\Dd^2\E^{(H)}+m \E^{(\Phi)}$ and then notice
\be \eqalign{
I=\Dd^2\E^{(H)}+m\E^{(\Phi)} ~=~ &\left(\g -b_1\right)m~\Dd^2\bar{\Phi}
~+~\left(2a_3-\g\right)m^2~\Dd^2H     \cr
&~+2a_2~\Dd^2{\rm D}^2H ~+~2b_2m^2~\Phi   ~~~.
} \ee
If we choose $\g=b_1=2a_3=2a_1,~a_2=0$ we obtain $I=2b_2m^2~\Phi$. 
Now we can follow two possible routes
\begin{enumerate}
\item $b_2\neq 0$:
$b_2$ can be anything besides zero and in that case on-shell $I=0\leadsto
\Phi=0$ we find all the desired constraints for $H$ and the action is
\be \eqalign{
S=\int d^8z &{\Big \{} c~H{\Dd}^{\gd}{\rm D}^2\Dd_{\gd}H
~+cm^2 \, H^2 
+2c m~H\left(\Phi+\bar{\Phi}\right)  \cr
&~+2c~\Phi\bar{\Phi} {\Big \}}
+\int d^6z~ b_2m~\Phi\Phi  ~~~.
}
\ee
\item $b_2=0$:
If we set $b_2$ to zero, then $I$ identically vanish. That means the $\Dd^2
\E^{(H)}+m\E^{(\Phi)}=0$ can be treated as a Bianchi identity and the 
corresponding action is invariant under a symmetry. The symmetry of the 
action that generates the above Bianchi identity is
\bea
\d_G H\sim \Dd^2 L+{\rm D}^2\bar{L} ~~, \IEEEyesnumber\\
\d_G \Phi\sim m\Dd^2 L  ~~~.
\eea
Due to this symmetry, the chiral superfield $\Phi$ can be gauged away 
completely and therefore it's equation of motion (or the Bianchi identity) 
will give the desired constraint of ${\rm D}^2H=0$. The action for this case 
is
\be \eqalign{ {~}
S=\int d^8z & {\Big \{ }c~H{\Dd}^{\gd}{\rm D}^2\Dd_{\gd}H 
~+cm^2\, H^2 ~+2c m~H\left(\Phi+\bar{\Phi}\right) 
~+~ 2c~\Phi\bar{\Phi}{\Big \}}  ~~~,
} \ee
and the gauge fixed action is identical with the action obtained from 
the first derivation. We would like to know if similar `St\"{u}ckelberg' 
constructions can occur for the higher superspin theories, like it is the 
case for the higher spin theories
\end{enumerate}

\section{New Massive $Y=3/2$ Theory}
\label{3/2}
~~Now we will follow a similar strategy to build a theory of superspin 
$\frac{3}{2}$. The starting point is the theory of superhelicity $\frac{3}{
2}$, in specific we choose the theory of non-minimal supergravity ($s
=1$ in \cite{H-IntSpinI}). Non-minimal supergravity is formulated in terms 
of $H_{\a\ad}$ and $\chi_{\a}$. We will add mass corrections to that 
action and check if 1) we can make $\chi_{\a}$ vanish on-shell 
(auxiliary status) and 2) we can generate the constraints on $H_{
\a\ad}$ demanded by representation theory ${\rm D}^\a H_{\a\ad}=0$,
~$\Box H_{\a\ad}=m^2 H_{\a\ad}$.  The starting action is given by
\be  \eqalign{
S=\int d^8z&\left\{\vphantom{\frac12} H^{\a\ad}{\rm D}^{\g}\Dd^2{\rm 
D}_{\g}H_{\a\ad}\right.~~~~~~\,~+a_1mH^{\a\ad}(\Dd_\ad\chi_\a-{\rm D}_\a
\bar{\chi}_\ad)\cr
&-2~H^{\a\ad}\Dd_{\ad}{\rm D}^2\chi_{\a}+c.c.~~+a_2mH^{\a\ad}
({\rm D}^2H_{\a\ad}+\Dd^2H_{\a\ad}) \cr
&-2~\chi^{\a}{\rm D}^2\chi_{\a}+c.c.~~~~~~~~+a_3m\chi^\a\chi_\a+c.c.\cr
&+2~\chi^{\a}{\rm D}_{\a}\Dd^{\ad}\bar{\chi}_{\ad}~~~~~~~~~~~~+a_4m^2H^{
\a\ad}H_{\a\ad} \vphantom{\frac12}{ \Big \} }  ~~~,
} \ee
and the equations of motion are:
\be \eqalign{  {~~~~~~~~}
\E^{(H)}_{\a\ad}&=2{\rm D}^\g\Dd^2{\rm D}_\g H_{\a\ad} +2({\rm 
D}_\a\Dd^2\bar{\chi}_\ad-\Dd_\ad{\rm D}^2\chi_\a)+a_1m(\Dd_\ad
\chi_\a-{\rm D}_\a\bar{\chi}_\ad)\cr
&~~~+2a_2m({\rm D}^2H_{\a\ad}+\Dd^2H_{\a\ad})+2a_4m^2H_{
\a\ad} ~~,
} \ee
\bea
\E^{(\chi)}_\a&=-4{\rm D}^2\chi_\a+2{\rm D}_{\a}\Dd^{\ad}\bar{\chi
}_{\ad}-2{\rm D}^2\Dd^\ad H_{\a\ad}+a_1m\Dd^\ad H_{\a\ad}+2a_3m\chi_\a 
~~~.
\eea

Now we may use these equations and attempt to remove any $H_{
\a\a}$-dependence to derive one equation that depends solely on 
$\chi_\a$. That will tell us if we can pick coefficients in a way that 
$\chi_\a$ vanishes on-shell. Consider the following linear combination 
of equations of motion where each such equation of motion is obtained 
by the variation of the action with regard to the respective superfields 
indicated by the subscripts in the first equation below:
\be \eqalign{
I_\a &=A{\rm D}^2\Dd^\ad \E^{(H)}_{\a\ad}+B{\rm D}^2\Dd^2\E^{(\chi
)}_{\a}+m^2\E^{(\chi)}_{\a}  {~~~~~~~~~~~~~~~~~~~~~~~}
 {~~~~~~~~~~~~~~~~~~~~~~~}  \cr
&=- 2\left(A + B\right)\Box{\rm D}^2\Dd^\ad H_{\a\ad} {~~~~}
~+ 2\left(A+B\right){\rm D}^2\Dd^2{\rm D}_{\a}\Dd^{\ad}\bar{\chi}_{\ad}{~~~}
~-Aa_1m{\rm D}^2\Dd^\ad{\rm D}_\a\bar{\chi}_\ad          \cr
&~~\,~+2 \left(Aa_4-1\right)m^2{\rm D}^2\Dd^{\ad}H_{\a\ad}{}~- 4\left(
A+ B\right)\Box{\rm D}^2\chi_{\a}
{~~~~~~~~~~~~}~-4m^2{\rm D}^2\chi_\a  \cr
&~~\,~+\left(a_1\right)m^3\Dd^\ad H_{\a\ad} {~~~~~~~~~~~~\,~}
~+ 2\left(Aa_1+ Ba_3\right)m{\rm D}^2\Dd^2\chi_{\a}
{~\,~} ~+2m^2{\rm D}_\a\Dd^\ad\bar{\chi}_{\ad} 
\cr
&{~~~~~~~~~~~~~~~~~~~~~~~~~~~~~~~~~~~~~~}  
{~~~~~~~~~~~~~~~~~~~~~~~~~~~~~~~~~~~~~~} 
~~\,~+2a_3m^3\chi_\a  ~~~.
} \ee

The following choice of coefficients will remove any $H_{\a\ad}$ 
dependence
from the equation above:
\bea
A+B=0~~,~~A a_4 - 1=0~~,~~a_1=0 ~~~, \IEEEyesnumber
\eea
and imposing these leads to the form of $I_\a$ to be given by
\be
I_\a ~=-4m^2{\rm D}^2\chi_\a+2Ba_3m{\rm D}^2\Dd^2\chi_{\a}
~+2m^2{\rm D}_\a\Dd^\ad\bar{\chi}_{\ad}+2a_3m^3\chi_\a    ~~~.
\ee

From this we see there is no choice of coefficients that will make 
$\chi_\a$ vanish on-shell. Therefore we must introduce an auxiliary 
superfield. Its purpose will be to impose a constraint on $\chi_\a$ 
when it vanishes. That constraint will be used to simplify the above 
expression for $I_\a$ and set $\chi_\a$ to zero. But a more careful 
examination of $I_\a$ will convince us that there is no unique 
constraint on $\chi_\a$ that will make all terms (except the last 
one) vanish. The inescapable conclusion is that we have to treat 
$\chi_\a=0$ as the desired constraint. This suggests that we must 
introduce a spinorial superfield $u_\a$ that couples with $\chi_\a$ 
through only a mass term $\sim mu^\a\chi_\a$. Hence when 
$u_\a=0$ then immediately we see $\chi_\a=0$.

We must update the action with the introduction of a few new terms:
the interaction term $m u^\a\chi_\a$ and the kinetic energy terms for 
$u_\a$ (the most general quadratic action). The new action is
\be \eqalign{
S=\int d^8z {\Big \{}\vphantom{\frac12} &H^{\a\ad}{\rm D}^{\g}\Dd^2{\rm 
D}_{\g}H_{\a\ad}
~~~~~~~~~~~~~~~~~~~~~~~~~~~~~~~~~~~~~~
~~~~~+\gamma m u^\a\chi_\a +c.c.  \cr
&-2~H^{\a\ad}\Dd_{\ad}{\rm D}^2\chi_{\a}+c.c.~~+a_2mH^{\a\ad}
{\rm D}^2 H_{\a\ad}+c.c.~+b_1 u^\a{\rm D}^2 u_\a+c.c.\cr  
&-2~\chi^{\a}{\rm D}^2\chi_{\a}+c.c.~~~~~~~~+a_3m\chi^\a\chi_\a
+c.c. ~~~~~~~~~+ b_2u^\a\Dd^2 u_\a+c.c.\cr
&+2~\chi^{\a}{\rm D}_{\a}\Dd^{\ad}\bar{\chi}_{\ad}~~~~~~~~~~~~+
a_4m^2 H^{\a\ad} H_{\a\ad}~~~~~~~~~~~+b_3 u^\a\Dd^\ad{\rm 
D}_\a\bar{u}_\ad \cr
&~~~~~~~~~~~~~~~~~~~~~~~~~~~~~~~~~~~~~~~
~~~~~~~~~~~~~~~~~~~~~~~~~~+b_4u^\a{\rm D}_\a\Dd^\ad\bar{
u}_\ad \cr
&~~~~~~~~~~~~~~~~~~~~~~~~~~~~~~~~~~~~~~~
~~~~~~~~~~~~~~~~~~~~~~~~~~+b_5m u^\a u_\a \vphantom{\frac12}
{\Big \}} ~~~, 
}\ee
and the updated equations of motion are
\be \eqalign{ {~~}
\E^{(H)}_{\a\ad}&=2{\rm D}^\g\Dd^2{\rm D}_\g H_{\a\ad} +2({\rm D}_\a
\Dd^2\bar{ \chi}_\ad-\Dd_\ad{\rm D}^2\chi_\a)+2a_2m({\rm D}^2H_{\a
\ad}+\Dd^2H_{\a \ad})   \cr
&~~~~+2a_4m^2H_{\a\ad}  ~~, }
\ee
\be
\E^{(\chi)}_\a=-4{\rm D}^2\chi_\a+2{\rm D}_{\a}\Dd^{\ad}\bar{\chi}_{\ad
}-2{\rm D}^2\Dd^\ad H_{\a\ad}+2a_3m\chi_\a+\gamma m u_\a  {~~~~,
~~~~~~~~~~~~~~~}
\ee
\be
\E^{(u)}_\a=2b_1{\rm D}^2 u_\a+2b_2\Dd^2 u_\a+b_3\Dd^\ad{\rm D}_\a
\bar{u}_\ad+b_4{\rm D}_\a\Dd^\ad\bar{u}_\ad+2b_5m u_\a+\gamma 
m\chi_\a {~~~.~~\,}
\ee

Now we repeat the process of eliminating $H_{\a\ad}$, but since $u_\a$ 
doesn't couple to $H_{\a\ad}$ nothing will be changed regarding the $H_{
\a\ad}$-dependent terms. The same choice of coefficients as in $(24)$ must 
be made to remove $H_{\a\ad}$. So the updated expression for $I_\a$ is
\be \eqalign{
I_\a
=&2Ba_3m{\rm D}^2\Dd^2\chi_{\a} ~~-4m^2{\rm D}^2\chi_\a     \cr
&+B\gamma m{\rm D}^2\Dd^2 u_\a ~+2m^2{\rm D}_\a\Dd^\ad\bar{\chi}_{\ad}\cr
&+\gamma m^3 u_\a ~~~~~~~~\,~+2a_3m^3\chi_\a  ~~~.
} \ee

Now we want to use the equation of motion of $u_\a$ to remove any 
dependences on $\chi_\a$ in order to derive an equation of $u_\a$.
For that we calculate the updated version of $I_{\a}$ which we denote
by $J_{\a}$ whose explicit form is given by
\be \eqalign{
J_\a ~&=~ I_\a+m K{\rm D}^2\E^{(u)}_\a+m\Lambda{\rm D}_\a\Dd^\ad\bar{\E
}^{(u)}_\ad  \cr
&=[2Ba_3]{\rm D}^2\Dd^2\chi_{\a}~~~~~~~\,~~~~~+[B\gamma+2Kb_2+
\Lambda b_3] m{\rm D}^2\Dd^\ad u_\a  \cr
&~~~~-[4-K\gamma]m^2{\rm D}^2\chi_\a~~~~\,~+[Kb_3+2\Lambda b_2]
m{\rm D}^2\Dd^\ad{\rm D}_\a \bar{u}_\ad \cr
&~~~~+[2+\Lambda\gamma]m^2{\rm D}_\a\Dd^\ad\bar{\chi}_\ad~~+
 [\Lambda (2b_4-b_3)]{\rm D}_\a\Dd^2{\rm D}^\b u_\b  \cr
&~~~~+[2a_3]m^3\chi_\a~~~~~~~~~~~~~~\,+\gamma m^3 u_\a   \cr
&~~~~+[Kb_5]m^2{\rm D}^2 u_\a~~~~~~~~~~+[\Lambda b_5]m^2{\rm 
D}_\a\Dd^\ad\bar{u}_\ad ~~~.
} \ee
If we choose
\bea
a_3=0~~,~~-4+K\gamma=0~~,~~2+\Lambda\gamma=0 ~~~,
\eea
we derive an equation of motion for $u_\a$ in the form
\be \eqalign{
J_\a
=&[B\gamma+2Kb_2+\Lambda b_3]m{\rm D}^2\Dd^\ad u_\a~+[Kb_5]
m^2{\rm D}^2 u_\a  \cr
&+[Kb_3+2\Lambda b_2]m{\rm D}^2\Dd^\ad {\rm D}_\a \bar{u}_\ad~ +
[\Lambda b_5]m^2{\rm D}_\a\Dd^\ad\bar{u}_\ad \cr
&+[\Lambda (2b_4-b_3)]{\rm D}_\a\Dd^2{\rm D}^\b u_\b \cr
&+\gamma m^3 u_\a
} \ee
Now we are in position to choose coeffecients so as to make $u_\a$ 
vanish on-shell by selecting
\bea
B\gamma+2Kb_2+\Lambda b_3=0~,~Kb_3+2\Lambda b_2=0~,~2b_4-
b_3 =0~,~b_5=0~,~\gamma\neq 0
\eea

Since $u_\a=0$ on-shell, now we can reverse the arguments. Its equation 
of motion will give $\chi_\a=0$ and that will put constraints on $H_{\a\ad}$: 
${\rm D}^2\Dd^\ad H_{\a\ad}=0$
\be \eqalign{
\E^{(H)}_{\a\ad}&=2{\rm D}^\g\Dd^2{\rm D}_\g H_{\a\ad}+2a_2m({\rm D}^2
H_{\a\ad} +\Dd^2H_{\a\ad})+2a_4m^2H_{\a\ad} ~~~, \cr
\E^{(\chi)}_\a&=-2{\rm D}^2\Dd^\ad H_{\a\ad}  ~~~.
} \ee
Finally because of ${\rm D}^2\Dd^\ad H_{\a\ad}=0$ we see that
\be
{\rm D}^\a\E^{(H)}_{\a\ad}=2a_2m{\rm D}^\a\Dd^2 H_{\a\ad}+2a_4m^2{\rm 
D}^\a H_{\a\ad}  ~~~.
\ee
For $a_2=0,~a_4\neq 0$ this gives ${\rm D}^\a H_{\a\ad}=0$. Thus the equation 
of motion for $H_{\a\ad}$ becomes the Klein-Gordon equation with $a_4=1$
\be
\Box H_{\a\ad}=m^2 H_{\a\ad}
\ee
To complete the analysis we look for the consistency and non-trivial solution 
of the systems of equations (20), (28), (30), $a_2=0$, and $a_4=1$.  A solution 
exists and it is
\be \eqalign{
a_1=0~~&,~~b_1=\text{free, can be set to zero}~~,~~\gamma=1~~,~~
\Lambda=-2 ~~~,  \cr
a_2=0~~&,~~b_2=\frac{1}{6}~~,~~~~~~~~~~~~~~~~~~~~~~~~~~~~,
~~A=1~~~, \cr
a_3=0~~&,~~b_3=\frac{1}{6}~~,~~~~~~~~~~~~~~~~~~~~~~~~~~~~,
~~B=-1 ~~~, \cr
a_4=1~~&,~~b_4=\frac{1}{12}~~,~~~~~~~~~~~~~~~~~~~~~~~~~~~,
~~K=4 ~~~, \cr
&~~\,~b_5=0 ~~.
}  \ee

The final action takes the form
\be \eqalign{
S=\int d^8z& {\Big \{} \vphantom{\frac12} H^{\a\ad}{\rm D}^{\g}\Dd^2{\rm 
D}_{\g}H_{\a\ad} ~~~~~~~~~~+
m u^\a\chi_\a +c.c. \cr
&~~-2~H^{\a\ad}\Dd_{\ad}{\rm D}^2\chi_{\a}+c.c.~+\frac{1}{6}u^\a\Dd^2 u_\a
+c.c.  \cr
&~~-2~\chi^{\a}{\rm D}^2\chi_{\a}+c.c.~~~~~~~+\frac{1}{6} u^\a\Dd^\ad
{\rm D}_\a\bar{u}_\ad    \cr
&~~+2~\chi^{\a}{\rm D}_{\a}\Dd^{\ad}\bar{\chi}_{\ad}~~~~~~~~~~~+
\frac{1}{12} u^\a{\rm D}_\a \Dd^\ad\bar{u}_\ad \cr
&~~~~~~~~+ m^2H^{\a\ad}H_{\a\ad} \vphantom{\frac12} {\Big \} }  ~~~.
} \ee

This is the superspace action that describes a superspin $Y=\frac{3}{2}$ 
system with the minimum number of auxiliary superfields and has a 
massless limit that gives non-minimal supergravity. This action is a 
representative of a family of actions that are all equivalent and connected 
through superfields redefinitions of the form
\bea
\chi_a \rightarrow\chi_\a +z_1 u_\a +w_1\Dd^\ad H_{\a\ad}\\
u_\a \rightarrow u_\a +z_2\chi_\a +w_2\Dd^\ad H_{\a\ad}, ~\text{where} 
~z_i,w_i ~\text{are complex}
\eea

\section{Summary}
~~ We started with the $\frac{3}{2}$ superhelicity theory of non-minimal supergravity, formulated in terms of a real vector superfield $H_{\a\ad}$ and a fermionic compensator $\chi_\a$. We then added mass terms to it in an attempt to discover a theory for massive superspin $\frac{3}{2}$ system, only to find that it is not possible and we need the help of an extra fermionic auxiliary superfield $u_\a$ which must couple only to $\chi_\a$ through a mass term. Finally using the equations of motion we manage to show that on-shell $u_\a=0\leadsto\chi_\a=0\leadsto\D^\a H_{\a\ad}=0\leadsto\Box H_{\a\ad}=m^2 H_{\a\ad}$

We have managed to derive yet another formulation of massive supergravity and most importantly probe the set of auxiliary superfields required for the construction of higher superspin theories.

\section*{Acknowledgements}
 This research has been supported in part by NSF Grant  
 PHY-09-68854, the J.~S. Toll Professorship endowment 
 and the UMCP Center for String \& Particle Theory.

\bibliography{Bibtex}

\end{document}